\documentclass[pre,english,preprint]{revtex4}
\usepackage{graphicx,amssymb,epsfig,babel}
\usepackage{mathrsfs}
 
\begin{document}  
 
\title{Graph Compression \\ --- Save Information by Exploiting Redundancy}
\author{Jie Sun, Erik M. Bollt}
\affiliation{Department of Mathematics \& Computer Science, Clarkson University, Potsdam, NY 13699-5815}
\author{Daniel ben-Avraham}
\affiliation{Department of Physics, Clarkson University, Potsdam, NY 13699-5820}

\begin{abstract}
In this paper we raise the question of how to compress sparse graphs. By introducing the idea of redundancy, we find a way to measure the overlap of neighbors between nodes in networks. We exploit symmetry and information by making use of the overlap in neighbors and analyzing how information is reduced by shrinking the network and using the specific data structure we created, we generalize the problem of compression as an optimization problem on the possible choices of orbits. To find a reasonably good solution to this problem we use a greedy algorithm to determine the orbit of symmetry identifications, to achieve compression. Some example implementations of our algorithm are illustrated and analyzed.
\end{abstract}

\maketitle

\newpage

\section{Introduction}
Complex networks have been studied extensively in  recent years in the fields of mathematics, physics, computer science, biology, sociology, etc \cite{WSnature,WSsw,BAscience,BAreview}. Various networks are used to model and analyze real world objects and their interactions with each other. For example, in sociology, airports and airflights that connect them can be represented by a network \cite{Pajek}; in biology, yeast reactions is also modeled by network \cite{SUNyeast}; etc. The mathematical terminology for a network is conveniently described in the language of graph theory. A common encoding of graphs uses an adjacency matrix, or an edge list, when the adjacency matrix is sparse \cite{ADJACENCYMATRIX}. However, even for a large network the edge list contains a large information storage. In the case that some important network is transfered frequently between computers, it will save time and cost if there is a scheme to efficiently encode, and therefore compress the network first. Fundamentally we find it a relevant issue to ask how much information is necessary to present a given network, and how symmetry can be exploited to this end. 

In this paper we will demonstrate one way to reduce the information storage of a network by using the idea that habitually  graphs have many nodes that share many common neighbors.  So instead of recording all the links we could rather just store some of them and the difference between neighbors. The ideal compression ratio using this scheme will be $\eta=\frac{2}{<k>}$ where $<k>$ is the average degree of the network, compared to the standard compression using Yale Sparse Matrix Format \cite{YSMF1,YSMF2} which gives $\eta_{Y}=\frac{1}{2}+\frac{1}{<k>}$. In practice this ratio is not attainable but the real compression ratio is still better than using YSMF as shown by our results.

A graph $G=(V,E)$ is a set of vertices (or nodes) $V=\{v_{1},v_{2},...,v_{N}\}$ together with edges (or links) $E=\{(v_{i},v_{j})\}$ which are the connected pairs. Graphs are often used to model networks. It is sometimes convenient to call the vertices that connect to a vertex $i$ in a graph to be the neighbors of $i$. We will only consider undirected and unweighted graph in this paper.
\begin{figure}[htbp]
\includegraphics[height=1.10in,width=1.29in]{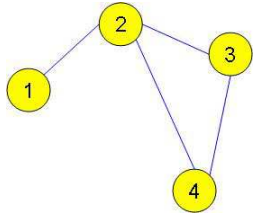}
\caption{A drawing of a planar embedding of an example graph.}
\label{figNetwork}
\end{figure}

A drawing as in Figure \ref{figNetwork} allows us to directly visualize the graph (i.e. the nodes and the connections between them), but 
a truism that anyone who works with real world graphs from real data knows is that commonly those graphs are so large that even a drawing will not give any insight.  Visualizing structure in graphs of such sizes ($N > 100$ to $1000$) begs for some computer assistance.

An {\it Adjacency matrix} is a common, although inefficient data representation of a graph. The adjacency matrix $A_{G}$ of a graph $G=(V,E)$ is a $N \times N$ square matrix where $N$ is the number of vertices of the graph and the entries $a_{i,j}$ of $A_{G}$ are defined by:
\begin{eqnarray}\label{adjancymat}
a_{i,j} &=& 1 \mbox{  if node i and node j are connected}\nonumber\\
a_{i,j} &=& 0 \mbox{  else}
\end{eqnarray}

For example, the adjacency matrix $A_{G}$ for the graph in Figure \ref{figNetwork} is
\begin{equation}\label{netmat}
A_{G}=
\left[
\begin{array}{ccccccc}
0 & \mbox{ } & 1 & \mbox{ } & 0 & \mbox{ } & 0 \\
1 & \mbox{ } & 0 & \mbox{ } & 1 & \mbox{ } & 1 \\
0 & \mbox{ } & 1 & \mbox{ } & 0 & \mbox{ } & 1 \\
0 & \mbox{ } & 1 & \mbox{ } & 1 & \mbox{ } & 0
\end{array}
\right].
\end{equation}

However, in the case that the number of edges in a graph are so few that the corresponding adjacency matrix is sparse, the {\it edge list} will be used instead. The edge list is a list of all the pairs of nodes that form edges in a graph. It is essentially the same as the edge set $E$ for a graph $G=(V,E)$. Using edge list $E_{G}$ to represent the same graph as above we will have:
\begin{equation}\label{netedges}
E_{G}=
\{\{1,2\},\{2,3\},\{3,4\},\{2,4\}\}.
\end{equation}

Note here that in the edge list we actually record the label of nodes for each edge in the graph, so for undirected graph, we can exchange the order for each pair of nodes.

We will only consider sparse simple graphs, whose adjacency matrices will thus be binary sparse matrices, and the standard information storage for such graphs or matrices will be the information units that are needed for the corresponding edge list (or two dimensional arrays).

We now sharpen the definition for the {\it unit of information} in our context. From the perspective of information theory, a message which contains $N$ different symbols will require $log_{2}N$ bits for each symbol, without any further coding scheme. The edge list representation is one example of a text file which contains $N$ different symbols (often represented by natural numbers from $1$ to $N$) for a graph containing $N$ vertices. Note that the unit of information depends only on the number of symbols that appear in the message, i.e. the number of vertices in a graph, so for any given graph this will be a fixed number. Thus, when we restrict the disscussion to any particular graph, it is convenient to assume that each pair of labels in the edge list requires one information unit without making explicit what is the size of that unit. For example, the above graph requires $4$ information units. In this paper we will focus on how to represent the same graph using fewer information units than its original representation.

\section{A Motivating Example and the Idea of Redundancy}
As a motivating example, let us consider the following graph and its edge list.

\begin{figure}[htbp]
\includegraphics[height=1.00in,width=3.52in]{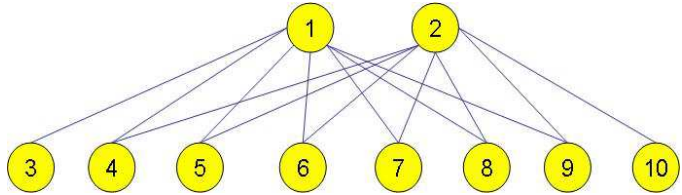}
\caption{An extreme example which shows similarity between vertices.}
\label{figP}
\end{figure}

Note that here the neighbors of node $1$ are almost the same as those of node $2$. The edge list $E_{G}$ for this graph will be: 
\begin{eqnarray}\label{exedges}
E=
\{\{1,3\},\{1,4\},\{1,5\},\{1,6\},\{1,7\},\{1,8\},\{1,9\},\nonumber\\
\{2,4\},\{2,5\},\{2,6\},\{2,7\},\{2,8\},\{2,9\},\{2,10\}\}.
\end{eqnarray}

This requires $14$ information units for the edge list. However, if we look back to the graph, we note that in this graph there are many common neighbors between node $1$ and node $2$, so there is a great deal of information redundancy. Considering the subgraphs, the neighbors of node $1$ are almost the same as the neighbors of node $2$, except that node $3$ links to $1$, but not $2$, while node $10$ links to $2$, but not $1$. 

\begin{figure}[htbp]
\includegraphics[height=0.90in,width=2.80in]{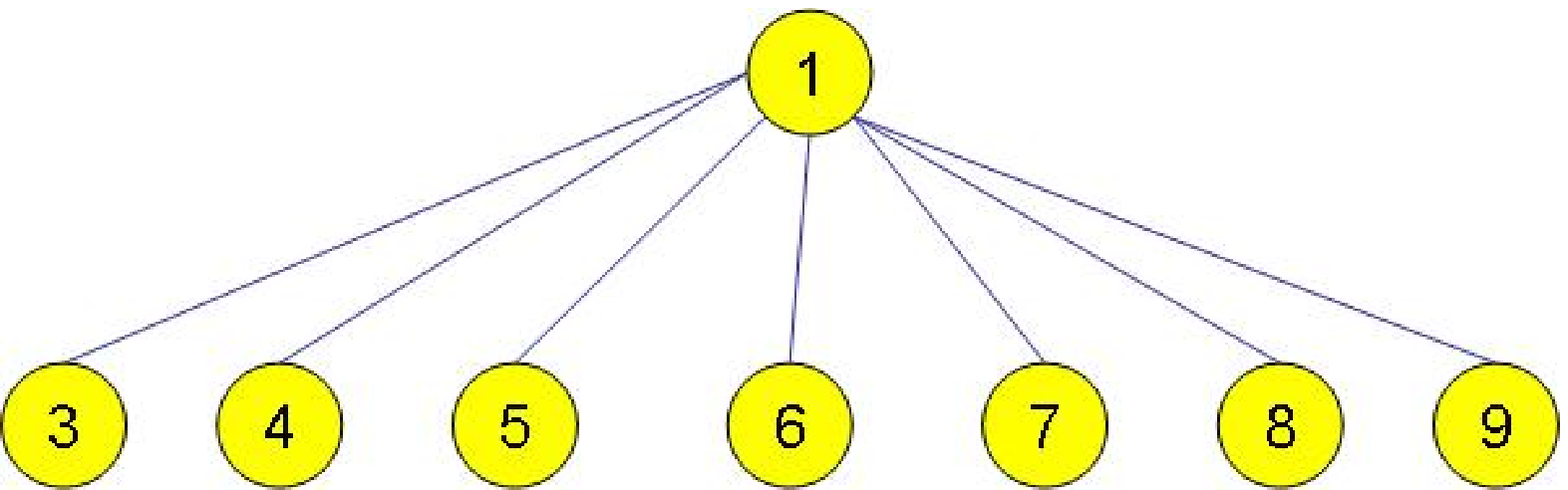}
\hspace{0.50in}
\includegraphics[height=0.90in,width=2.80in]{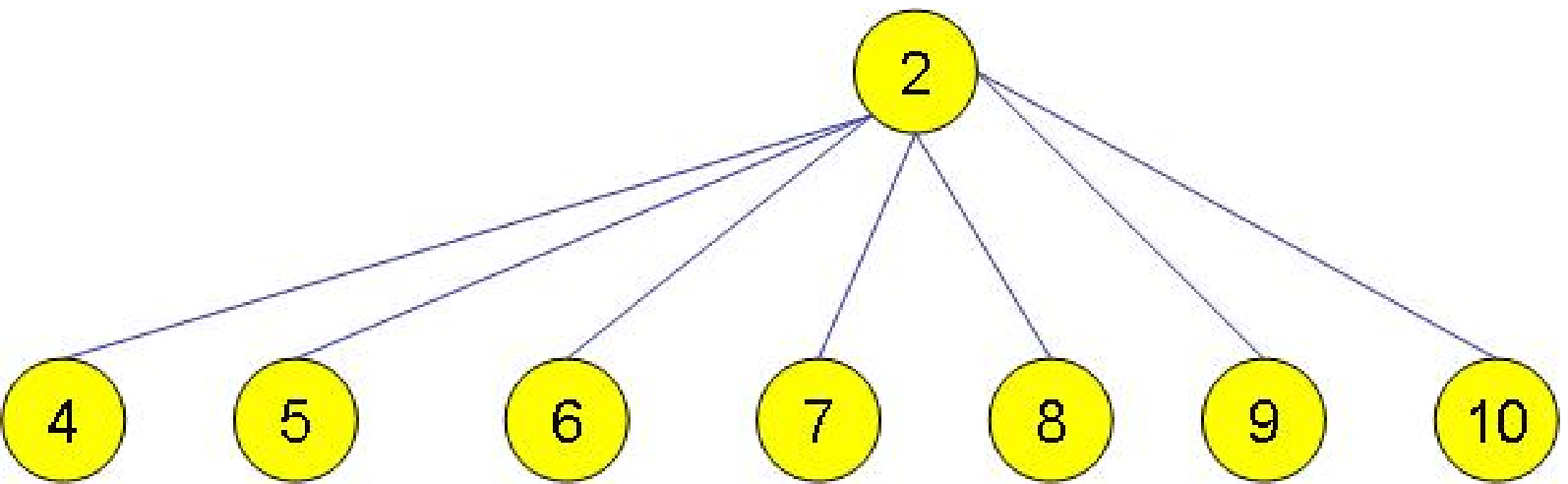}
\caption{Similar subgraphs of the original graph. Here the subgraph containing node 1 (on the left) is very similar to the one dominated by node 2 (on the right).}
\label{subgraphs}
\end{figure}

Taking the redundancy into account, we generate a new way to describe the same graph, exploiting the graphs. In the graph of Figure \ref{figP}, we see that the subgraph including vertices $1,3,4,5,6,7,8,9$ is very similar to the subgraph including vertices $2,4,5,6,7,8,9,10$, see Figure \ref{subgraphs}. We exploit this redundancy in our coding.

We store the subgraph which only consists of node 1, and all its neighbors.  Then, we add just two more parameters,
\begin{equation}\label{alphaset}
 \alpha=(1,2)
\end{equation}
and 
\begin{equation}\label{betaset}
 \beta=\{-3,10\}
\end{equation}
that allows us to reconstruct the original graph. Here the ordered pair $\alpha=(1,2)$ tells us that in order to reconstruct the original graph we need to first copy node $1$ to node $2$. By copy, we mean the addition of a new node into the exsiting graph with label $2$, and then linking all the neighbors of node $1$ to the new node $2$. 

\begin{figure}[htbp]
\includegraphics[height=1.91in,width=4.22in]{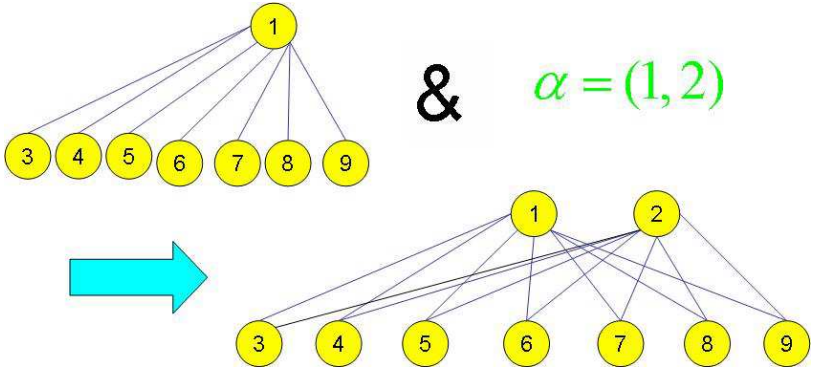}
\caption{Construct from the subgraph and parameter $\alpha=(1,2)$. 'Copy' from node $1$ to node $2$.}
\label{subgraphalpha}
\end{figure}

The set $\beta=\{-3,10\}$ tells us that we should then delete the link that connects the new node 2 and 3 and add a new link between 2 and 10.

\begin{figure}[htbp]
\includegraphics[height=2in,width=5in]{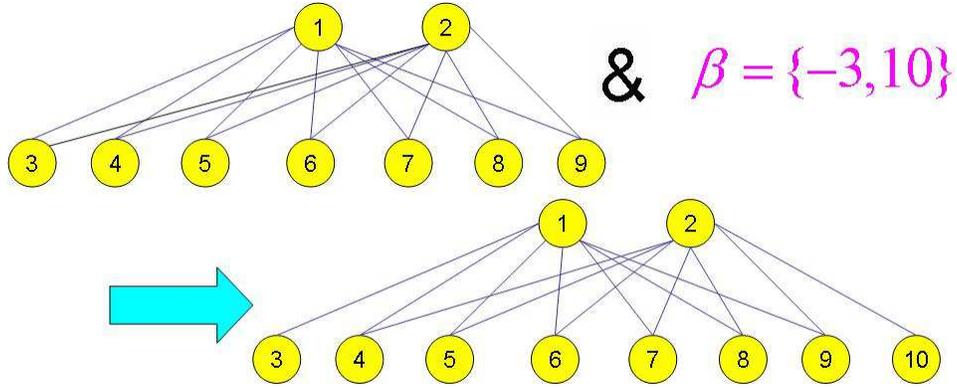}
\caption{Add and delete links according to $\beta=\{-3,10\}$.}
\label{subgraphbeta}
\end{figure}

After all these operations we see that we successfully reconstruct the graph with fewer information units, in this case, nearly half as many as the original edge list. So instead of equation (\ref{exedges}), we may use the edge list of the subgraph
\begin{equation}\label{subgraphedges}
E_{SG}=
\{\{1,3\},\{1,4\},\{1,5\},\{1,6\},\{1,7\},\{1,8\},\{1,9\}\}
\end{equation}
as well as two sets
\begin{eqnarray}\label{alphabeta}
 \alpha=(1,2)\nonumber\\
 \beta=\{-3,10\}
\end{eqnarray}
to represent the same graph.

\begin{figure}[htbp]
\includegraphics[height=2.20in,width=4.20in]{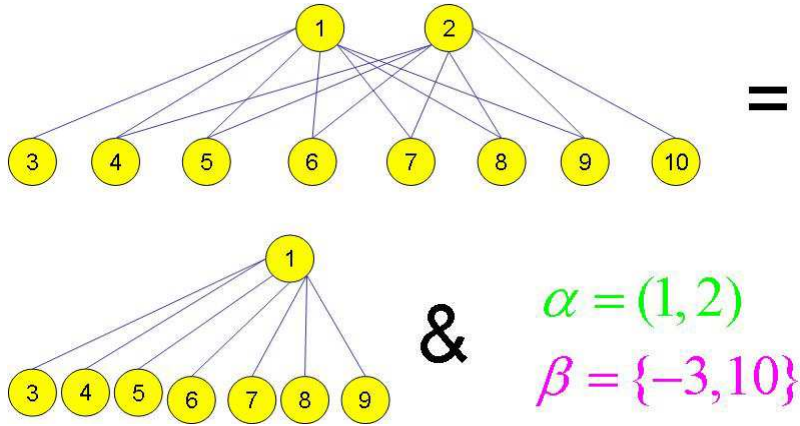}
\caption{Reconstruction of the original graph using a subgraph and the parameters $\alpha$ and $\beta$.}
\label{graphrec}
\end{figure}

The above example suggests that by exploiting symmetry of the graph, we might be able to reduce the information storage for certain graphs by using a small subgraph as well as $\alpha$ and $\beta$ as defined above.

However, there remains the question of how to choose the pair of vertices so that we actually reduce the information, and which is the best possible pair? It is important to answer these questions since most of the graphs are so large that we never will be able to see the symmetry just by inspection as we did for the above toy example. 

In the following we answer the first question, and partly the second, by using a greedy algorithm. In  section 3 we will define information redundancy for a binary sparse matrix and show that it reveals the neighbor similarity between vertices in a graph which is represented by its corresponding adjacency matrix. Then in section 4 we will give a detailed description of our algorithm which allows us to implement our main idea.  Then in section 5 we will show some examples of these applications followed by  discussion in section 6.

\section{Information Redundancy and Compression of Sparse Matrices}
\subsection{How to Choose Pairs of Vertices to Reduce Information}
The graphs we seek to compress are typically represented by large sparse adjacency matrices. An edge-list is a specific data structure for representing such matrices, to reduce information storage. We will consider the edge-list form to be the standard way of storing sparse matrices, which requires $M$ units of information for a graph with $M$ edges. There are approaches of compressing sparse matrices, among which the most general is the Yale Sparse Matrix Format \cite{YSMF1,YSMF2}, which does not make any assumption on the structure of the matrix and only requires $\frac{1}{2}(M+N)$ units of information. There are other approaches, such as \cite{TARJANYAO} which emphasize not only the storage but also the cost for data access time. We will focus on the data storage, so the Yale Format will be considered as a basic benchmark approach for compression of a sparse matrix, to which we will compare our results. The Yale format yields the compression ratio:
\begin{equation}\label{YaleFormat}
\eta_{Y}=\frac{M+N}{2M}=\frac{1}{2}+\frac{1}{<k>}
\end{equation}
where $<k>=\frac{2M}{N}$ is the average degree of the graph.

We will show our approach of compressing the sparse matrices by first illustrating how the redundancy of a binary sparse matrix will be defined regarding to our specific operation on the matrix.

Generally, the adjacency matrix is a binary sparse matrix, $A=\{a_{ij}\}$ where $a_{ij}$ equals $0$ or $1$ indicating the connectivity between node $i$ and $j$. For a simple graph consisting of $M$ edges this matrix has $2M$ nonzero entries, but since it is symmetric only half of them are necessary to represent the graph, which yields $M$ units of information for the edge-list.

Now, if two nodes $i$ and $j$ in the graph share a lot of similar neighbors, in the adjacency matrix row $i$ and row $j$ will have a lot of common column entries, and likewise for column $i$ and column $j$ (due to the symmetry of the matrix). 

Suppose that we apply the operation  to the graph, mentioned in the last section, by choosing $\alpha=(i,j)$ and the corresponding $\beta$, we will not need row $j$ and column $j$ in the matrix, to represent the graph. The number of nonzero entries in row $j$ and column $j$ is $2k_{j}$ where $k_{j}$ is the degree of node $j$ in the graph. By doing that, the number of nonzero entries in the new adjacency matrix becomes $2M-2k_{j}$, which requires $M-k_{j}$ units of information. However, the extra information we have to record is encoded in $\alpha$ and $\beta$. $\alpha$ always has two entries, which requires $1$ unit of information, and the units of information for $\beta$ depend on the number of different neighbors between node $i$ and node $j$. If $i$ and $j$ have $\Delta_{ij}$ different neighbors, the size of $\beta$ will be 
\begin{equation}\label{sizebeta}
|\beta|=\Delta_{ij},
\end{equation}
and the units of information for $\beta$ will thus be $\frac{1}{2}\Delta_{ij}$. Taking both the reduction of the matrix and the extra information into account, the actual information it requires after the operation is
\begin{equation}\label{actinfo}
M-k_{j}+1+\frac{1}{2}\Delta_{ij}=M-(k_{j}-1-\frac{1}{2}\Delta_{ij}). 
\end{equation}
This is true for $i$ different from $j$. We could extend the operation to allow 
\begin{equation}\label{selfalpha}
\alpha=(i,i), 
\end{equation}
meaning a self-match, then we will put all the neighbors of $i$ into the corresponding set $\beta$, and then delete these links associated with $i$. Then by a similar argument we find that after this operation we need
\begin{equation}\label{newinfo}
M-k_{i}+1+\frac{1}{2}k_{i}=M-(\frac{1}{2}k_{i}-1)
\end{equation} 
units of information using the new format.

Note that here we need to clarify exactly the meaning of different neighbors since in the case that $i$ and $j$ are connected $i$ is a neighbor of $j$ but $j$ is not, and likewise for $j$. However, this extra information can be simply encoded in $\alpha$ by making the following rule: $\alpha=(i,j)$ means when we reconstruct we do not connect $i$ and $j$ and $\alpha=(i,-j)$ means we connect $i$ and $j$ when we reconstruct. Then we can write $\Delta_{ij}=\|A(i,:)-A(j,:)\|_{1}-2a_{ij}$. 

From the above discussion we see that if we define
\begin{eqnarray}\label{reddef}
r_{ij} &=& k_{j}-1-\frac{1}{2}\Delta_{ij}\mbox{,}\mbox{ }\mbox{ }\mbox{ }\mbox{ }i \neq j\nonumber\\
r_{ii} &=& \frac{1}{2}k_{i}-1\mbox{ }\mbox{ }\mbox{ }\mbox{ }\mbox{ }
\end{eqnarray}
then by choosing $\alpha=(i,j)$, $r_{ij}$ measures exactly the amount of information it reduces. We call $r_{ij}$ the information redundancy between node $i$ and $j$. Note here that in general this redundancy is not symmetric in $i$ and $j$, since for any pair of nodes $\Delta_{ij}$ is symmetric but the degree of these two nodes can be different, and deleting the node with higher degree will always reduce more units of information compared to deleting the lower degree node.

We form the redundancy matrix $R$ by setting the entry in row $i$ and columnn $j$ to be $r_{ij}$. We perform the shrinking operation for the pair with maximum $r_{ij}$, thus saving the maximum amount of information.

For example, again using  the graph from section $2$, the adjacency matrix is:

\begin{equation}\label{matex}
A =
\left[
\begin{array}{ccccccccccccccccccc}
0 & \mbox{ } & 0 & \mbox{ } & 1 & \mbox{ } & 1 & \mbox{ } & 1 & \mbox{ } & 1 & \mbox{ } & 1 & \mbox{ } & 1 & \mbox{ } & 1 & \mbox{ } & 0\\
0 & \mbox{ } & 0 & \mbox{ } & 0 & \mbox{ } & 1 & \mbox{ } & 1 & \mbox{ } & 1 & \mbox{ } & 1 & \mbox{ } & 1 & \mbox{ } & 1 & \mbox{ } & 1\\
1 & \mbox{ } & 0 & \mbox{ } & 0 & \mbox{ } & 0 & \mbox{ } & 0 & \mbox{ } & 0 & \mbox{ } & 0 & \mbox{ } & 0 & \mbox{ } & 0 & \mbox{ } & 0\\
1 & \mbox{ } & 1 & \mbox{ } & 0 & \mbox{ } & 0 & \mbox{ } & 0 & \mbox{ } & 0 & \mbox{ } & 0 & \mbox{ } & 0 & \mbox{ } & 0 & \mbox{ } & 0\\
1 & \mbox{ } & 1 & \mbox{ } & 0 & \mbox{ } & 0 & \mbox{ } & 0 & \mbox{ } & 0 & \mbox{ } & 0 & \mbox{ } & 0 & \mbox{ } & 0 & \mbox{ } & 0\\
1 & \mbox{ } & 1 & \mbox{ } & 0 & \mbox{ } & 0 & \mbox{ } & 0 & \mbox{ } & 0 & \mbox{ } & 0 & \mbox{ } & 0 & \mbox{ } & 0 & \mbox{ } & 0\\
1 & \mbox{ } & 1 & \mbox{ } & 0 & \mbox{ } & 0 & \mbox{ } & 0 & \mbox{ } & 0 & \mbox{ } & 0 & \mbox{ } & 0 & \mbox{ } & 0 & \mbox{ } & 0\\
1 & \mbox{ } & 1 & \mbox{ } & 0 & \mbox{ } & 0 & \mbox{ } & 0 & \mbox{ } & 0 & \mbox{ } & 0 & \mbox{ } & 0 & \mbox{ } & 0 & \mbox{ } & 0\\
1 & \mbox{ } & 1 & \mbox{ } & 0 & \mbox{ } & 0 & \mbox{ } & 0 & \mbox{ } & 0 & \mbox{ } & 0 & \mbox{ } & 0 & \mbox{ } & 0 & \mbox{ } & 0\\
0 & \mbox{ } & 1 & \mbox{ } & 0 & \mbox{ } & 0 & \mbox{ } & 0 & \mbox{ } & 0 & \mbox{ } & 0 & \mbox{ } & 0 & \mbox{ } & 0 & \mbox{ } & 0\\
\end{array}
\right],
\end{equation}

and the corresponding redundancy matrix is:
\begin{equation}\label{redatex}
R =
\left[
\begin{array}{ccccccccccccccccccc}
2.5 & \mbox{ } & 5 & \mbox{ } & -3 & \mbox{ } & -2.5 & \mbox{ } & -2.5 & \mbox{ } & -2.5 & \mbox{ } & -2.5 & \mbox{ } & -2.5 & \mbox{ } & -2.5 & \mbox{ } & -4\\
5 & \mbox{ } & 2.5 & \mbox{ } & -4 & \mbox{ } & -2.5 & \mbox{ } & -2.5 & \mbox{ } & -2.5 & \mbox{ } & -2.5 & \mbox{ } & -2.5 & \mbox{ } & -2.5 & \mbox{ } & -3\\
3 & \mbox{ } & 2 & \mbox{ } & -0.5 & \mbox{ } & 0.5 & \mbox{ } & 0.5 & \mbox{ } & 0.5 & \mbox{ } & 0.5 & \mbox{ } & 0.5 & \mbox{ } & 0.5 & \mbox{ } & -1\\
2.5 & \mbox{ } & 2.5 & \mbox{ } & -0.5 & \mbox{ } & 0 & \mbox{ } & 1 & \mbox{ } & 1 & \mbox{ } & 1 & \mbox{ } & 1 & \mbox{ } & 1 & \mbox{ } & -0.5\\
2.5 & \mbox{ } & 2.5 & \mbox{ } & -0.5 & \mbox{ } & 1 & \mbox{ } & 0 & \mbox{ } & 1 & \mbox{ } & 1 & \mbox{ } & 1 & \mbox{ } & 1 & \mbox{ } & -0.5\\
2.5 & \mbox{ } & 2.5 & \mbox{ } & -0.5 & \mbox{ } & 1 & \mbox{ } & 1 & \mbox{ } & 0 & \mbox{ } & 1 & \mbox{ } & 1 & \mbox{ } & 1 & \mbox{ } & -0.5\\
2.5 & \mbox{ } & 2.5 & \mbox{ } & -0.5 & \mbox{ } & 1 & \mbox{ } & 1 & \mbox{ } & 1 & \mbox{ } & 0 & \mbox{ } & 1 & \mbox{ } & 1 & \mbox{ } & -0.5\\
2.5 & \mbox{ } & 2.5 & \mbox{ } & -0.5 & \mbox{ } & 1 & \mbox{ } & 1 & \mbox{ } & 1 & \mbox{ } & 1 & \mbox{ } & 0 & \mbox{ } & 1 & \mbox{ } & -0.5\\
2.5 & \mbox{ } & 2.5 & \mbox{ } & -0.5 & \mbox{ } & 1 & \mbox{ } & 1 & \mbox{ } & 1 & \mbox{ } & 1 & \mbox{ } & 1 & \mbox{ } & 0 & \mbox{ } & -0.5\\
2 & \mbox{ } & 3 & \mbox{ } & -1 & \mbox{ } & 0.5 & \mbox{ } & 0.5 & \mbox{ } & 0.5 & \mbox{ } & 0.5 & \mbox{ } & 0.5 & \mbox{ } & 0.5 & \mbox{ } & -0.5\\
\end{array}
\right],
\end{equation}

The maximum entry in $R$ is $r_{12}=r_{21}=5$, indicating that either choice of $\alpha=(1,2)$ or $\alpha=(2,1)$ will give the maximum information reduction, and the corresponding $\beta$ can be obtained by recording the column entries in row $1$ and row $2$ according to our rule.

In the above discussion we only consider a one step shrinking operation on the graph and find out the direct relationship between the maximum information reduction and the redanduncy matrix. But we know that after deleting one node the resulting graph is still sparse and so could be compressed further by our scheme. The question is then how to successively choose $\alpha$ and $\beta$ to obtain the best overall compression.

\subsection{On Greedy Optimization of The $\alpha,\beta$, Orbit}
Let $\alpha_{t}=(i_{t},j_{t})$ denote the operation at step $t$, $t=1,2,...,T$ (here the sign for $j_{t}$ would not affect our analysis so by convience we just write $j_{t}$). In order to analyze the multi-step effect, we first consider how the adjacency matrix $A$ is affected by the orbit $\{\alpha_{t}\}$. Let $A_{0}=A$ be the original adjacency matrix. Let $A_{t}$ be the corresponding adjacency matrix after applying $\alpha_{t}$ and the entries in it be $A_{t}(i,j)$. On deleting node $j_{t}$ we actually set row and column $j_{t}$ to be zero in $A_{t-1}$ and all the other entries are unchanged, to obtain the new matrix $A_{t}$, i.e.

\begin{eqnarray}\label{updateA}
A_{t}(i,j) &=& A_{t-1}(i,j)\mbox{ }\mbox{ }\mbox{ }\mbox{ }\mbox{ }\mbox{ if }i,j \neq j_{t}\nonumber\\
A_{t}(i,j) &=& 0 \mbox{ }\mbox{ }\mbox{ }\mbox{ }\mbox{ }\mbox{ if }i = j_{t} \mbox{ or }j = j_{t}.
\end{eqnarray}

So by induction we see that

\begin{eqnarray}\label{updateA2}
A_{t}(i,j) &=& A_{0}(i,j)\mbox{ }\mbox{ }\mbox{ }\mbox{ }\mbox{ }\mbox{ if }i,j \notin \{j_{1},...,j_{t}\}\nonumber\\
A_{t}(i,j) &=& 0 \mbox{ }\mbox{ }\mbox{ }\mbox{ }\mbox{ }\mbox{ if }i \in \{j_{1},...,j_{t}\}\mbox{ or }j \in \{j_{1},...,j_{t}\}.
\end{eqnarray}

Then we analyze how the redundancy matrix $R$ changes. Use $R_{t}$ to represent the redundancy matrix, $k_{t}(i)$ the degree of node $i$, and $\Delta_{t}(i,j)$ the number of different neighbors of node $i$ and $j$, associated with the graph of $A_{t}$. Since our goal is to achieve compression, once a node is deleted in the graph it is useless for future operations. So we will set $R_{t}(i,j)=0$ if $i$ or $j$ has been deleted before, i.e.
\begin{equation}\label{newrij}
R_{t}(i,j)=0 \mbox{ }\mbox{ }\mbox{ }\mbox{ }\mbox{ }\mbox{ if }i \in \{j_{1},...,j_{t}\}\mbox{ or }j \in \{j_{1},...,j_{t}\}.
\end{equation}

Now for those $i$ and $j$ that have not been deleted, i.e. $i,j \notin \{j_{1},...,j_{t}\}$, by equation \ref{reddef} we see that $R_{t}(i,j)=k_{t}(j)-1-\frac{1}{2}\Delta_{t}(i,j)$ for $i \neq j$ and $R_{t}(i,i)=\frac{1}{2}k_{t}(i)-1$. Since $A_{t}$ is obtained by deleting row and columnn $j_{t}$ in $A_{t-1}$, the degree of each node changes according to:
\begin{eqnarray}\label{updatek}
k_{t}(i) &=& k_{t-1}(i)-A_{t-1}(i,j_{t})
\end{eqnarray}
and $\Delta_{ij}$ changes according to
\begin{eqnarray}\label{updatediff}
\Delta_{t}(i,j) &=& \Delta_{t-1}(i,j)-|A_{t-1}(i,j_{t})-A_{t-1}(j,j_{t})|
\end{eqnarray}

Thus, we conclude that for $i \neq j$
\begin{eqnarray}\label{redatt}
R_{t}(i,j) &=& k_{t-1}(j)-A_{t-1}(j,j_{t})-1-\frac{1}{2}[\Delta_{t-1}(i,j)-|A_{t-1}(i,j_{t})-A_{t-1}(j,j_{t})|]\nonumber\\
           &=& k_{t-1}(j)-1-\frac{1}{2}\Delta_{t-1}(i,j)-A_{t-1}(j,j_{t})+\frac{1}{2}|A_{t-1}(i,j_{t})-A_{t-1}(j,j_{t})|\nonumber\\
           &=& R_{t-1}(i,j)+[\frac{1}{2}|A_{t-1}(i,j_{t})-A_{t-1}(j,j_{t})|-A_{t-1}(j,j_{t})]
\end{eqnarray}
and for $i = j$
\begin{eqnarray}\label{red2att}
R_{t}(i,i) &=& \frac{1}{2}k_{t}(i)-1\nonumber\\
           &=& \frac{1}{2}(k_{t-1}(i)-A_{t-1}(i,j_{t}))-1\nonumber\\
           &=& R_{t-1}(i,i)-\frac{1}{2}A_{t-1}(i,j_{t}).
\end{eqnarray}

By induction, we obtain that for $i \neq j$:
\begin{eqnarray}\label{redattgen}
R_{t}(i,j) &=& R_{0}(i,j)+\sum_{\tau=1}^{t}[\frac{1}{2}|A_{\tau-1}(i,j_{\tau})-A_{\tau-1}(j,j_{\tau})|-A_{\tau-1}(j,j_{\tau})]
\end{eqnarray}
and for $i = j$:
\begin{eqnarray}\label{redattgen2}
R_{t}(i,i) &=& R_{0}(i,i)+\sum_{\tau=1}^{t}[-\frac{1}{2}A_{\tau-1}(i,j_{\tau})].
\end{eqnarray}

By use oft the fact that $i,j \notin \{j_{1},...,j_{t}\}$, by equation \ref{updateA}, we can simplify the above two expressions to yield,
\begin{eqnarray}\label{redattsim}
R_{t}(i,j) &=& R_{0}(i,j)+\sum_{\tau=1}^{t}[\frac{1}{2}|A_{0}(i,j_{\tau})-A_{0}(j,j_{\tau})|-A_{0}(j,j_{\tau})]\mbox{ }\mbox{ }\mbox{ }\mbox{ }\mbox{ }\mbox{ if }i \neq j\nonumber\\
R_{t}(i,i) &=& R_{0}(i,i)+\sum_{\tau=1}^{t}[-\frac{1}{2}A_{0}(i,j_{\tau})].
\end{eqnarray}

Note that if we choose a pair $(i_{t},j_{t})$ at step $t$, the information we save is measured by $R_{t-1}(i_{t},j_{t})$. Thus, for any orbit $\{\alpha_{t}=(i_{t},j_{t})\}_{t=1}^{T}$ satisfying $i_{t},j_{t} \notin \{j_{1},...,j_{t-1}\}$ for $t=2,3,...,T$ (we call such an orbit a {\it natural orbit}), the total information reduction (or information saving) will be:
\begin{eqnarray}\label{suminfosaving}
s(\{\alpha_{t}\}_{t=1}^{T}) &=& \sum_{t=1}^{T}R_{t-1}(i_{t},j_{t})\nonumber\\
                            &=& \sum_{t=1}^{T}[R_{0}(i_{t},j_{t})+c(i_{t},j_{t},t)]
\end{eqnarray}
where $c$ is defined by:
\begin{eqnarray}\label{defc}
c(i,j,t) &=& \sum_{\tau=1}^{t}[\frac{1}{2}|A_{0}(i,j_{\tau})-A_{0}(j,j_{\tau})|-A_{0}(j,j_{\tau})]\mbox{ }\mbox{ }\mbox{ }\mbox{ }\mbox{ }\mbox{ if }i \neq j\nonumber\\
c(i,i,t) &=& \sum_{\tau=1}^{t}[-\frac{1}{2}A_{0}(i,j_{\tau})].
\end{eqnarray}

So the compression problem can be stated as:
\begin{equation}\label{maxorbit}
\mbox{Find}\mbox{  }\max_{\{\alpha_{t}\}_{t=1}^{T}}s(\{\alpha_{t}\}_{t=1}^{T}). 
\end{equation}

One more thing to mention is that the length of the orbit, $T$, is also a variable, which could not be larger than $N$ since there are only $N$ nodes in the graph and it is meaningless to delete an `empty' node which does not even exist.

\section{Greedy Algorithm for Compression}
From the previous section we see that for a given adjacency matrix, the final compression ratio depends on the orbit $\{\alpha_{t}\}_{t=1}^{T}$ we choose, and the compression problem becomes an optimization problem. However, to find the maximum of $s$ and the corresponding best orbit is not trivial. One reason is that the number of natural orbits is of order $N!$, which makes it impractical to test and try for all possible orbits. Another reason which is crucial here is that for any given orbit of length $T$, evaluating $s$ costs $O(T^2)$ operations, making it hard to find an appropriate scheme to search for the true maximum or even the approximate maximum. Instead, we use a greedy algorithm to find an orbit which gives a reasonable compression ratio, and which is easy to apply.

The idea of the greedy algorithm is that at each iteration step we choose the pair of nodes $i_{t}$ and $j_{t}$ which maximizes $R_{t-1}(i,j)$ over all possible pairs, and we stop if the maximum value is non-positive. Also we need to record $\alpha$ and $\beta$ according to the graph.

Here we summarize the greedy algorithm as pseudocode:

Given the adjacency matrix $A$ of a graph ($N$ nodes and $M$ edges).

Begin: 

Set $A_{0}=A$;

Calculate $R_{0}(i,j)$ for all $i,j=1,...,N$. This forms the redundancy matrix $R_{0}=R$.

Set t=1.
\begin{enumerate}
\item \mbox{Let}\mbox{ }$R_{t-1}(i_{t},j_{t})$\mbox{ }\mbox{be the largest element in $R_{t-1}$}.\\
      \mbox{If}\mbox{ }$R_{t-1}(i_{t},j_{t})>0$\\
      \mbox{ }\mbox{record $\alpha_{t}=(i_{t},j_{t})$,}\\
      \mbox{ }\mbox{then go to step 2.}\\
      \mbox{Else,}\\
      \mbox{ }\mbox{End.}
\item 
      \mbox{Set}\mbox{ }$\beta_{t}$\mbox{ }\mbox{according to the difference between the two rows of}
      \mbox{ }$\alpha_{t}$\mbox{ }\mbox{in $A_{t-1}$,}\\
      \mbox{Update $A_{t-1}$ to $A_{t}$ according to (\ref{updateA})};\\
      \mbox{Update $R_{t-1}$ to $R_{t}$ according to (\ref{redatt}) and (\ref{red2att}) for $i,j \neq j_{t}$};\\
      \mbox{Set $R_{t}(i,j) = 0$ for $i$ or $j$ $ = j_{t}$}.
\item Set $t=t+1$ and go to step 1.
\end{enumerate}

The compressed version of the matrix will consist of: the final matrix $A_{T}$, the orbit $(\alpha_{1},...,\alpha_{T})$ and the vectors $\{\beta_{1},...,\beta_{T}\}$, which will allow us to reconstruct $A=A_{0}$ and any intermediate matrix $A_{t}$ during the compression process.

\section{Examples of Application to Graphs}
In this section we will show some examples of our compression scheme on several networks. We begin with the lattice graph, which is expected to be readily compressible due to the high degree of overlapping between neighbors of nodes. As a secondary example,  we add some random alterations, and apply our method to the corresponding Watts-Strogatz network. Finally we show some results for real-world networks.

\subsection{A Simple Benchmark Example: Lattice Graph}
One of the most symmetric graphs is the lattice graph, a one-dimensional chain where each site is connected to $\frac{k}{2}$ nearest neighbors to its right and left. In this case $<k> = k$ represents the degree of each vertex in the lattice graph. The total number of nodes is $N >> <k>$, the corresponding adjacency matrix is sparse.

\begin{figure}[htbp]
\includegraphics[height=3in,width=4in]{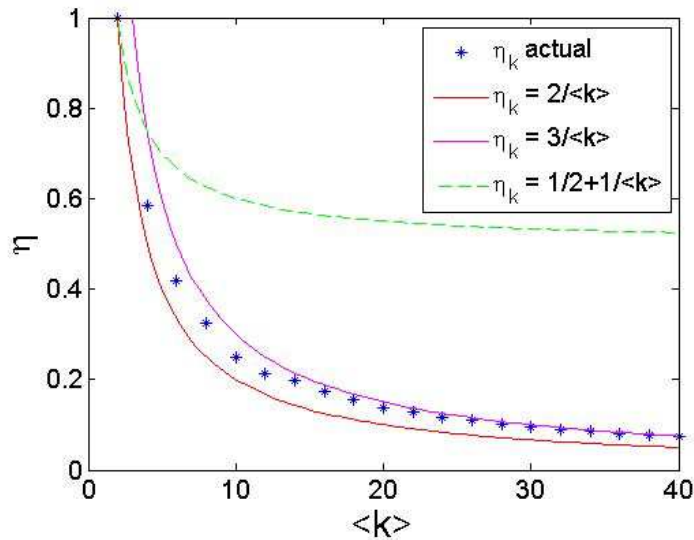}
\caption{Compression results for lattice graphs. Stars indicate the final compression ratios for the lattice graphs with $<k>$ $2$ to $40$. The compression limit is indicated by the bottom curve given by $\eta_{k} = \frac{2}{<k>}$, and we find that for $<k>$ large the compression ratio is close to the empirical formula: $\eta_{k} = \frac{3}{<k>}$ (upper curve). For comparison, we plot the result using YSMF (broken line): $\eta_{k} = \frac{1}{2} + \frac{1}{<k>}$. For $<k> > 2$, our algorithm always achieves a better result than the YSMF and the advantage increases with increasing $<k>$.}
\label{latticeresults}
\end{figure}

We implement our algorithm for lattice graph with different $<k>$. The results are shown in Figure \ref{latticeresults}. Here we take $N=500$.

\subsection{Compressing  a Watts-Strogatz Small-World Graph}
It is not surprising that the lattice graphs are easy to compress since these graphs are highly symmetric and nodes have lots of overlaps in their neighbors. However, in the case that we don't have such perfect symmetry, we still hope to achieve compression. Here we apply our algorithm to the WS graphs. The WS graph comes from the famous Watts-Strogatz model for real-world networks by showing the so called small-world phenomenon. The WS graph is generated from a lattice graph by the usual rewiring of each edge with some given probability $p$ from the uniform distribution.  

\begin{figure}[htbp]
\includegraphics[height=3in,width=4in]{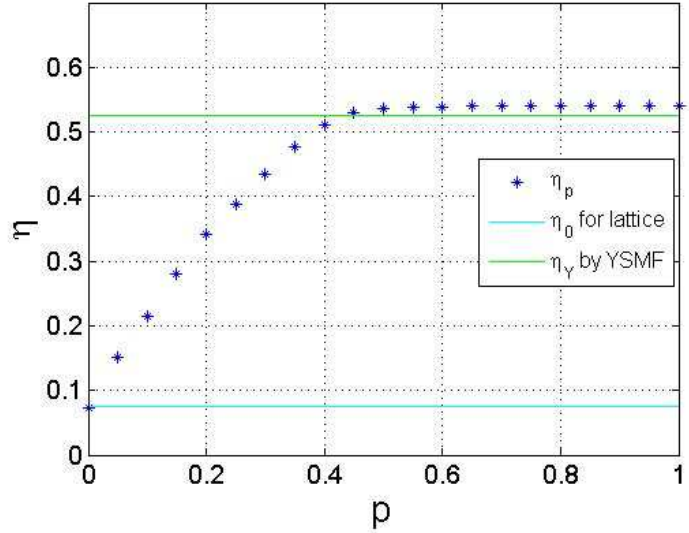}
\caption{Compression results for WS graphs. Here the base lattice graph is with $N=500$ and $<k>=40$. The stars show the compression results by our algorithm. The lower line is the compression ratio for the lattice $N=500$ and $<k>=40$ and the upper line is the ratio from the YSMF. We see that as $p$ increases there is less and less overlapping between neighbors in the network and the compression ratio increases. For $p\sim0.5$, we obtain worse result than YSMF.}
\label{wsresults}
\end{figure}

We apply our algorithm to WS graphs with different $p$ to explore how $p$ affects the compression behavior. Results are shown in Figure \ref{wsresults}.

\subsection{Real-World Graphs}
\begin{figure}[htbp]
\includegraphics[height=2.1in,width=2.8in]{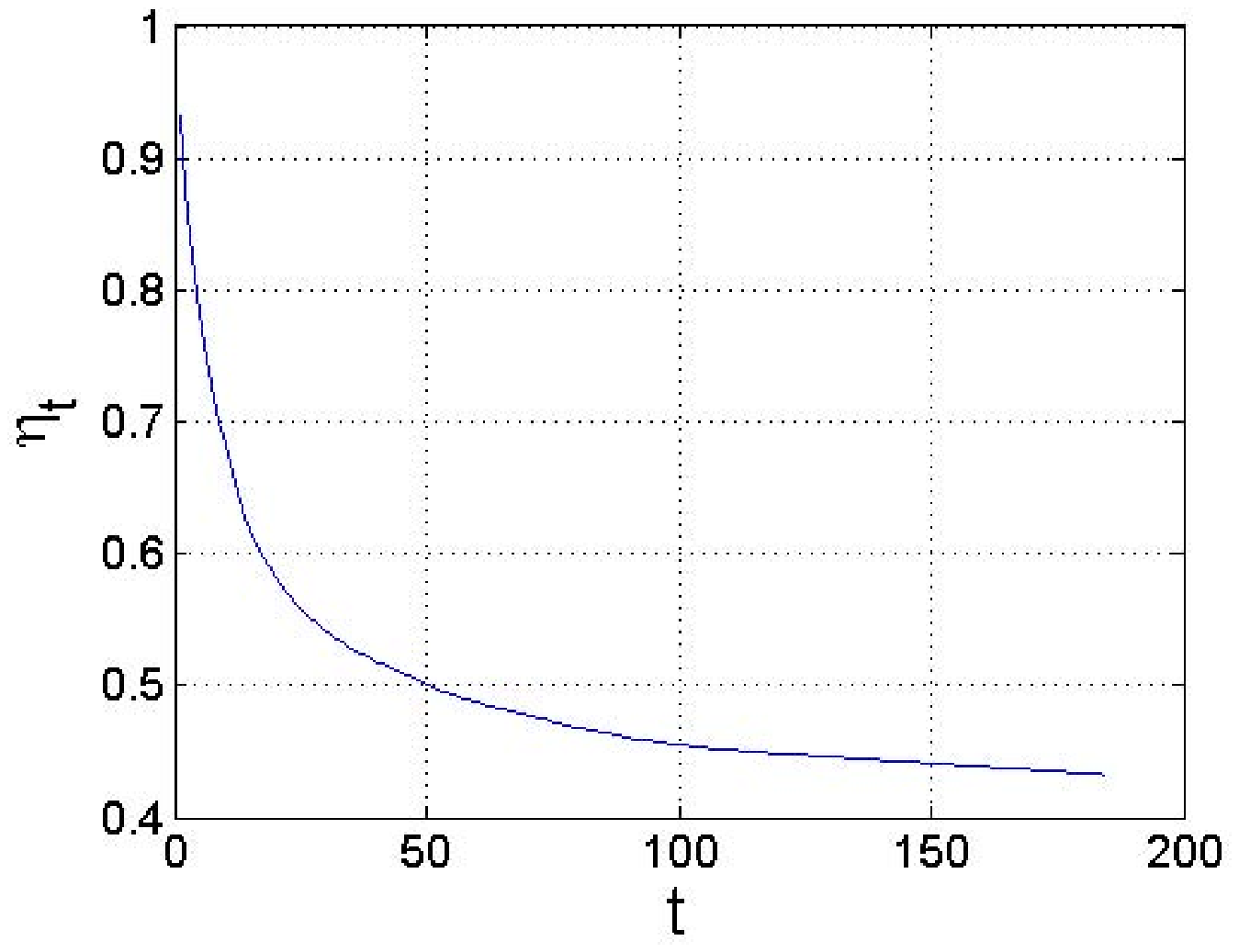}
\includegraphics[height=2.1in,width=2.8in]{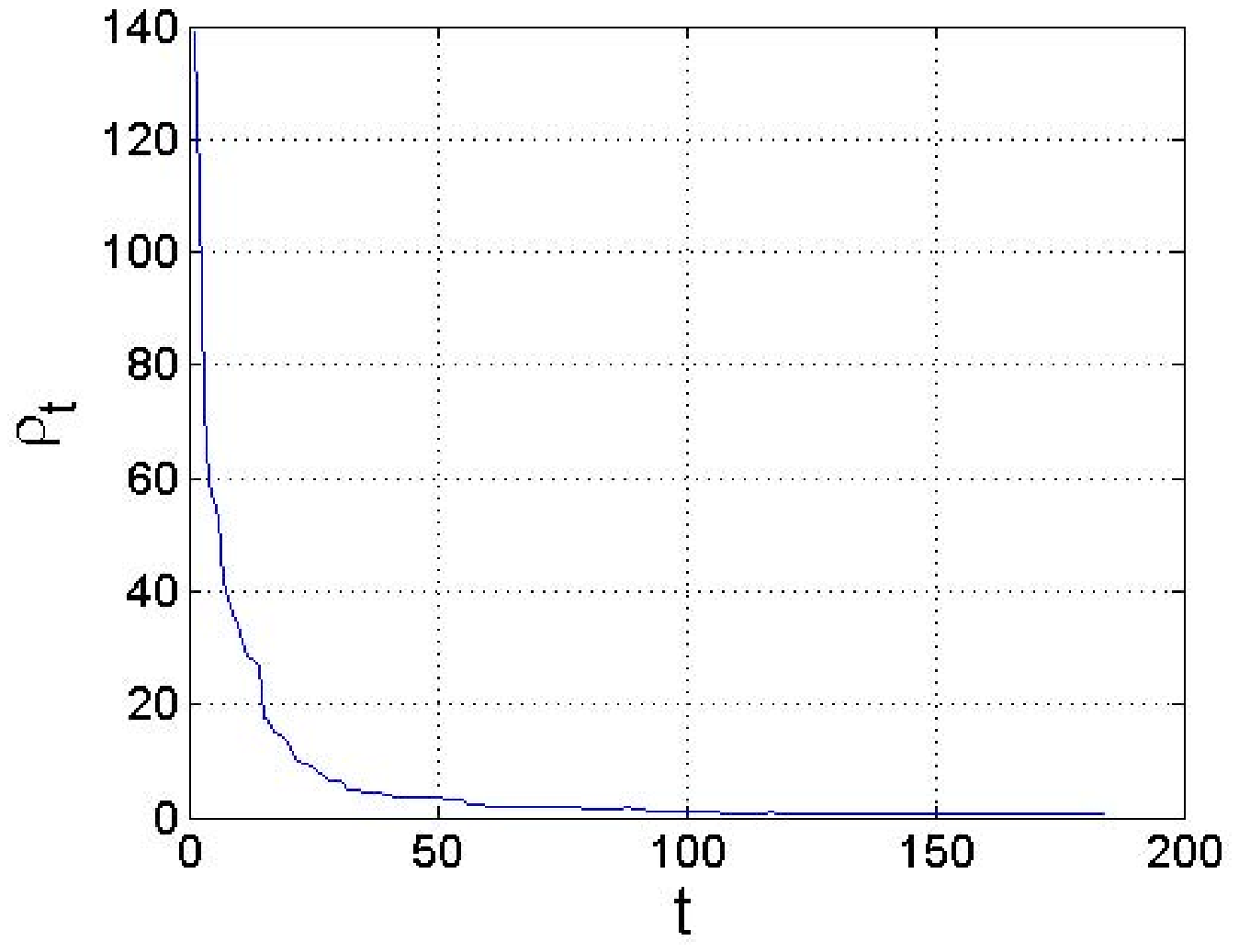}
\caption{Compression process for Metabolic network \cite{ARENASmetabolic}: compression ratio $\eta$ during each step (left), and information redundancy $\rho$ each step (right).}
\label{Metabolic}
\end{figure}

In the following we show the compression results for some real world graphs: a C.elegans metabolic network \cite{ARENASmetabolic} (Figure \ref{Metabolic}), a yeast network constructed from yeast reactions \cite{SUNyeast}, an email network \cite{ARENASemail}, and an airline network of flight connections \cite{Pajek}. In the table 1 we summarize the compression results for these real world graphs.

\begin{table}[htbp]
\begin{tabular}{|l|l|l|l|l|l|}
\hline
Network & N & $<k>$ & $\eta_{Y}$ & $\eta$ & $\eta_{*}$ \\ \hline
Lattice & N & $<k>$ & $\frac{1}{2}+\frac{1}{<k>}$ & $\frac{3}{<k>}$ & $\frac{2}{<k>}$ \\ \hline
Yeast \cite{SUNyeast} & 2361 & 6.08 & 0.66 & 0.50 & 0.33 \\ \hline
Metabolic \cite{ARENASmetabolic} & 453 & 9.01 & 0.61 & 0.43 & 0.22 \\ \hline
Email \cite{ARENASemail} & 1133 & 9.62 & 0.60 & 0.49 & 0.21 \\ \hline
Airline \cite{Pajek} & 332 & 12.81 & 0.58 & 0.31 & 0.16 \\ \hline
\end{tabular}
\caption{Compression results for some networks.}
\end{table}

\section{Discussion}
From the previous section we see that our algorithm works for various kinds of graphs and gives a reasonable result. The ideal limit of our method for a graph with $N$ nodes, $M$ edges and average degree $<k>=\frac{2M}{N}$, which is relative large, is $\frac{2}{<k>}$. This is obtained when each $\beta_{t}$ during the compression process is empty, meaning that most of the nodes share common neighbors, in which case we only need to record all the $\alpha_{t}$, requiring $\frac{N}{2}$ units of information and yields
\begin{equation}\label{idealcpr}
\eta = \frac{N}{2M} = \frac{2}{<k>}. 
\end{equation}

Notice that trees do not compress, since for trees $<k>=2$, so on average the overlap in neighbors will be even smaller (likely to be $0$), and a possible way to achieve compression is by self-matching for large degree nodes, for example, the hubs in a star graph. For comparison, the YSMF always gives the compression ratio
\begin{equation}\label{YSMFcpr}
\eta_{Y}=\frac{1}{2}+\frac{1}{<k>}
\end{equation}
which does not compress trees, and has a lower bound $\frac{1}{2}$, while our method in principle approaches $0$ as $<k> \rightarrow \infty$. Actually the compression ratio using YSMF can be achieved by choosing a special orbit in our approach which only contains self-matches $\alpha$, i.e.
\begin{equation}\label{selforbit}
\{\alpha_{t}\}_{t=1}^{T}=\{(i,i)\}_{i=1}^{N}.
\end{equation}
In this case the neighbors of each node will be put into corresponding $\beta$ sets and since any $\alpha_{i}$ contains the same pair of numbers $(i,i)$ we can just use one $i$ to represent the pair, resulting in a total $\frac{N+M}{2}$ information units. So our approach can be considered as a generalization of the YSMF.

However, as we observed in our compression results, the compression ratio given by \ref{idealcpr} is in general not attainable since it is only achieved for the ideal case that nearly every node in the graph shares the same neighbors, and yet the graph needs to be sparse! However, for lattices we observe that the actual compression ratio achieved by our algorithm is about $\frac{3}{<k>}$, which is of the same order as the ideal compression ratio. For WS graphs, when the noise $p$ is small, our algorithm achieves better compression ratio than YSMF, and the compression ratio is nearly linearly dependent on $p$ for $p<0.5$. For $p>0.5$ the graph resembles Erdos-Renyi random graphs \cite{ERgraph}, there is no symmetry between nodes to be used and thus our approach does not give good result, as compared to the YSMF.

For real world graphs, the results by our algorithm are better than using YSMF, but not as good as we observed for lattice graphs. This suggests that in real world graphs nodes, in general, share certain amount of common neighbors even when the total number of links is small. This kind of overlap in neighbors is certainly not as common as we see in lattice graphs since real world graphs in general have more complicated structures.

\section{Acknowledgements}
J.S. and E.M.B have been supported for this work  by the Army Research Office grant 51950-MA. E.M.B. has been further supported by the National Science Foundation under DMS-0708083 and DMS-0404778, and D.B.A is supported by the National Science Foundation under PHY-0555312. We thank Joseph D. Skufca and James P. Bagrow for discussion.

\end{document}